\documentclass[prd,letterpaper,superscriptaddress,twocolumn]{revtex4}%
\usepackage{graphicx}
\usepackage{bm}
\usepackage{latexsym}
\usepackage{epsf}
\usepackage{rotating}
\usepackage{epsfig,graphics,rotate,color}
\usepackage{wrapfig}
\usepackage{amssymb}
\usepackage{amsmath}
\usepackage{amsfonts}
\usepackage{subfigure}
\usepackage{array,hhline,dcolumn}%
\usepackage[normalem]{ulem}
\providecommand{\U}[1]{\protect\rule{.1in}{.1in}}
\begin{document}
\title{Measuring Active-to-Sterile Neutrino Oscillations with Neutral Current Coherent Neutrino-Nucleus Scattering}
\author{A.J. Anderson}
\affiliation{Massachusetts Institute of Technology, Cambridge, MA 02139, USA}
\author{J.M. Conrad}
\affiliation{Massachusetts Institute of Technology, Cambridge, MA 02139, USA}
\author{E. Figueroa-Feliciano}
\affiliation{Massachusetts Institute of Technology, Cambridge, MA 02139, USA}
\author{C. Ignarra}
\affiliation{Massachusetts Institute of Technology, Cambridge, MA 02139, USA}
\author{G. Karagiorgi}
\affiliation{Columbia University, New York, NY 10027, USA}
\author{K. Scholberg}
\affiliation{Duke University, Durham, NC 27708, USA}
\author{M.H. Shaevitz}
\affiliation{Columbia University, New York, NY 10027, USA}
\author{J. Spitz }
\affiliation{Massachusetts Institute of Technology, Cambridge, MA 02139, USA}

\begin{abstract} 
Light sterile neutrinos have been introduced as an explanation for a number of oscillation signals at $\Delta m^2 \sim 1$~eV$^2$. Neutrino oscillations at relatively short baselines provide a probe of these possible new states. This paper describes an accelerator-based experiment using neutral current coherent
neutrino-nucleus scattering to strictly search for active-to-sterile neutrino
oscillations. This experiment could, thus, definitively establish the existence of sterile neutrinos and provide constraints on their mixing parameters. A cyclotron-based proton beam can be directed to multiple targets, producing a low energy pion and muon decay-at-rest neutrino source with variable distance to a single detector. Two types of detectors are considered: a germanium-based detector inspired by the CDMS design and a liquid argon detector inspired by the proposed CLEAR experiment. 
\end{abstract}
\maketitle

\section{Introduction}
Sterile neutrino models have been invoked to explain a series of
intriguing oscillation signals at $\Delta m^2 \sim 1$~eV$^2$~\cite{Sorel:2003hf, Karagiorgi, Schwetz, Giunti}. These signals have relied on neutrino detection through charged current interactions. In the case
of charged current appearance, the signal is interpreted as an active flavor oscillating to another active flavor, which can occur at these high $\Delta m^2$ values if one or more neutrino mass states with $m_{4},...\sim1~\mathrm{eV}$ is added to the neutrino mass spectrum. The extra mass states are assumed to participate in neutrino oscillations, and must therefore be small admixtures of weakly-interacting neutrino flavor states, with the remaining flavor composition being {\it sterile} (i.e.~non-weakly-interacting). In the case of charged current disappearance,
the signal is interpreted as arising from active-flavor neutrino ($e,~\mu,~\tau$) oscillation to any other neutrino flavor ($e,~\mu,~\tau$, or $s$, with $s$ being the sterile flavor).

The oscillation probabilities for appearance and disappearance through charged current searches are expressed as functions of the active flavor content of the extra mass eigenstate(s) \cite{Sorel:2003hf,Karagiorgi}. In this paper, we assume that only one such extra mass state, $m_4$, exists. In that case, the oscillation probabilities are given by
\begin{equation}
\label{eq1}
P(\nu_{\alpha}\rightarrow\nu_{\beta\ne\alpha})=4|U_{\alpha4}|^2|U_{\beta4}|^2\sin^2(1.27\Delta m_{41}^2L/E)
\end{equation}
in the case of active appearance searches, and 
\begin{equation}
\label{eq2}
P(\nu_{\alpha}\rightarrow\nu_{\not{\alpha}})=4|U_{\alpha4}|^2(1-|U_{\alpha4}|^2)\sin^2(1.27\Delta m_{41}^2L/E)
\end{equation}
in the case of active disappearance searches, where $\alpha,\beta=e,\mu,\tau$; $\not{\alpha}$ corresponds to all flavors other than $\alpha$, including active and sterile; $|U_{\alpha4}|^2$ corresponds to the $\alpha$-flavor content of the fourth mass eigenstate; and $L$ and $E$ represent the neutrino travel distance and energy, respectively. Note that neither search case is purely sensitive to the sterile neutrino content of the extra neutrino mass state, $|U_{s4}|^2$. In this paper, we discuss a strictly neutral current search using coherent neutrino scattering that allows for pure active-to-sterile oscillation sensitivity. 

Coherent neutrino-nucleus scattering is a well-predicted neutral current
weak process with a high cross section in the standard model, as compared to other neutrino interactions at similar energies. Despite this, the coherent interaction has never been observed as the keV-scale nuclear recoil signature is difficult to detect. The newest generation of $\sim$10~keV threshold dark matter detectors provides sensitivity to coherent scattering~\cite{ourcoherent} as the interaction signal is nearly identical to that which is expected from WIMP interactions. 

An active-to-sterile neutrino oscillation search is motivated in Section~\ref{motiv}. We describe an experimental design which makes use of a high intensity pion- and muon- decay-at-rest (DAR) neutrino source in Section~\ref{neutsource}. The coherent scattering process is introduced and the experimental design is discussed in Section~\ref{coherentintro}. Sensitivities to neutrino oscillations at $\Delta m^2 \sim 1$~eV$^2$ are shown in Section~\ref{sensitivitysection}.

\section{Motivation for an Active-to-Sterile Oscillation Search}
\label{motiv}
A decade ago, sterile neutrino oscillation models were largely
motivated by the LSND anomaly
\cite{LSND,Peres:2000ic,Strumia:2002fw,Grimus:2001mn,Sorel:2003hf}.
This result presented a 3.8$\sigma$ excess of $\bar{\nu}_e$ events consistent with $\bar \nu_\mu
\rightarrow \bar \nu_e$ oscillations described by Eq.~\ref{eq1} at $\Delta m^2 \sim 1~{\rm
  eV}^2$ and $\sin^22\theta_{\mu e}=4|U_{e4}|^2|U_{\mu4}|^2\sim0.003$.
The apparent appearance signal is thus interpreted as indirect evidence for at least
one additional neutrino carrying the ability to mix with active flavors.  Being mostly sterile, an additional neutrino avoids conflict with measurements of the $Z$ invisible width~\cite{zwidth} (characteristic of three weakly-interacting light neutrino states) and the three-neutrino oscillation model established by solar~\cite{SuperKsolar, SNO, Kamland} and atmospheric/accelerator~\cite{SuperKatmos, SoudanII, k2kosc, MinosCC} experiments. 

The LSND signal was not present in a similar but less sensitive $\bar \nu_\mu \rightarrow \bar \nu_e$ oscillation search by the KARMEN experiment~\cite{KARMEN}. More recently, however, the MiniBooNE experiment~\cite{miniboone_nim} has explored the $\Delta m^2 \sim 1$~eV$^2$ parameter space and yielded a number of interesting results. MiniBooNE features a higher beam energy and larger distance than LSND but preserves the $L/E$ oscillation probability dependence, allowing for an independent cross check of the signal. In searching for $\nu_e$ appearance in a pure $\nu_\mu$ beam, MiniBooNE has excluded $\nu_\mu
\rightarrow \nu_e$ oscillations in the LSND $\Delta m^2$ range at the 90\%~CL~\cite{MBnu}. However, MiniBooNE's search for $\bar \nu_\mu \rightarrow \bar \nu_e$ oscillations in ``anti-neutrino-mode" is only consistent with the no oscillation hypothesis at the 0.5\% level~\cite{MBnubar}. The anti-neutrino result is consistent with LSND and $\bar \nu_\mu \rightarrow \bar \nu_e$ oscillations in the $\Delta m^2=0.1-1.0~\mathrm{eV}^2$ range. The statistics-limited measurement is expected to improve with additional data being taken through at least 2012.

Recently, further results for $\bar{\nu}_e$ disappearance at high
$\Delta m^2$ have been reported from short-baseline reactor anti-neutrino experiments. 
More specifically, a re-analysis of the anti-neutrino spectra produced by
fission products in a reactor core~\cite{Mueller} has led to an effect
termed ``the reactor anti-neutrino anomaly'', where the ratio of the observed anti-neutrino rate to the predicted rate deviates below unity at 98.6\%~CL~\cite{Lassere}. This can be interpreted as disappearance according to Eq.~\ref{eq2}, where charged current interactions of active flavors other than $e$ are kinematically forbidden, and/or where the oscillation was into a non-interacting sterile neutrino. Assuming CPT conservation, which requires that Eq.~\ref{eq2} holds for both neutrinos and anti-neutrinos, the strongest limits on $\bar{\nu}_e$ disappearance come from a joint analysis of KARMEN and LSND 
$\nu_e+{}^{12}{\rm C} \rightarrow {}^{12}{\rm N}_{gs} + e^-$ scattering events,
analyzed for evidence of $\nu_e$ disappearance \cite{nuedis}. The reactor-anomaly signal is found to be marginally consistent with the KARMEN and LSND $\nu_e$ disappearance results. 

The above experiments feature a single source, single detector design.
An alternative approach is a near-far detector configuration, where the
measured flux in the near detector replaces the first-principles flux prediction. A near-far design removes a significant source of uncertainty due to the flux prediction, especially if the detectors are built to be nearly identical. 
Using the near-far technique, the CDHS~\cite{CDHS}, CCFR~\cite{CCFR}, and SciBooNE/MiniBooNE~\cite{mbsbdis} experiments have probed neutrino disappearance at $\Delta m^2\sim1~\rm{eV}^2$ using $\nu_\mu$ charged current interactions. Among the recent near-far comparison data sets, the MINOS experiment has set the only limits on
active-to-sterile oscillations using neutral current interactions~\cite{MinosNC}. The resulting limits using both charged current and neutral current interactions present a challenge in fitting sterile neutrino oscillation models~\cite{georgiadpf}. 

The aforementioned results underscore the experimental and theoretical need for acquiring further data in addressing the possibility of sterile neutrinos~\cite{snacinprep}. 

\section{The Neutrino Source}
\label{neutsource}
A DAR neutrino source can be employed to search for active-to-sterile neutrino oscillations through the neutral current coherent scattering interaction. DAR neutrinos have
been identified as an excellent source for neutrino-nucleus coherent
scattering studies~\cite{scholberg,clear,Scholberg:2009ha} because the neutrinos are
produced in an energy region ($<$52.8 MeV) where
the coherent neutrino scattering cross section is higher than all others by about one order of magnitude.

A search for active-to-sterile oscillations is envisioned with a series of
measurements at different values of $L$ from the DAR source. In our
design, a cyclotron directs a proton beam to two graphite targets
embedded in a single iron shield. As the DAR neutrino flavor content and energy distribution is driven by the weak interaction, the well understood flux emitted isotropically from each target will be effectively identical, barring oscillations, at each baseline $L$.

As discussed in Ref.~\cite{ourcoherent}, an ideal neutrino interaction target for a DAR source is a direct dark matter detection device sensitive to keV-scale nuclear recoils. We consider two dark matter detector technologies; a germanium-based CDMS-style detector~\cite{CDMS} and a liquid argon-based one similar to the CLEAN~\cite{MiniClean} and CLEAR~\cite{Scholberg:2009ha} designs.

\subsection{About decay-at-rest sources}
\begin{figure}[b]\begin{center}
{\includegraphics[width=3.1in]{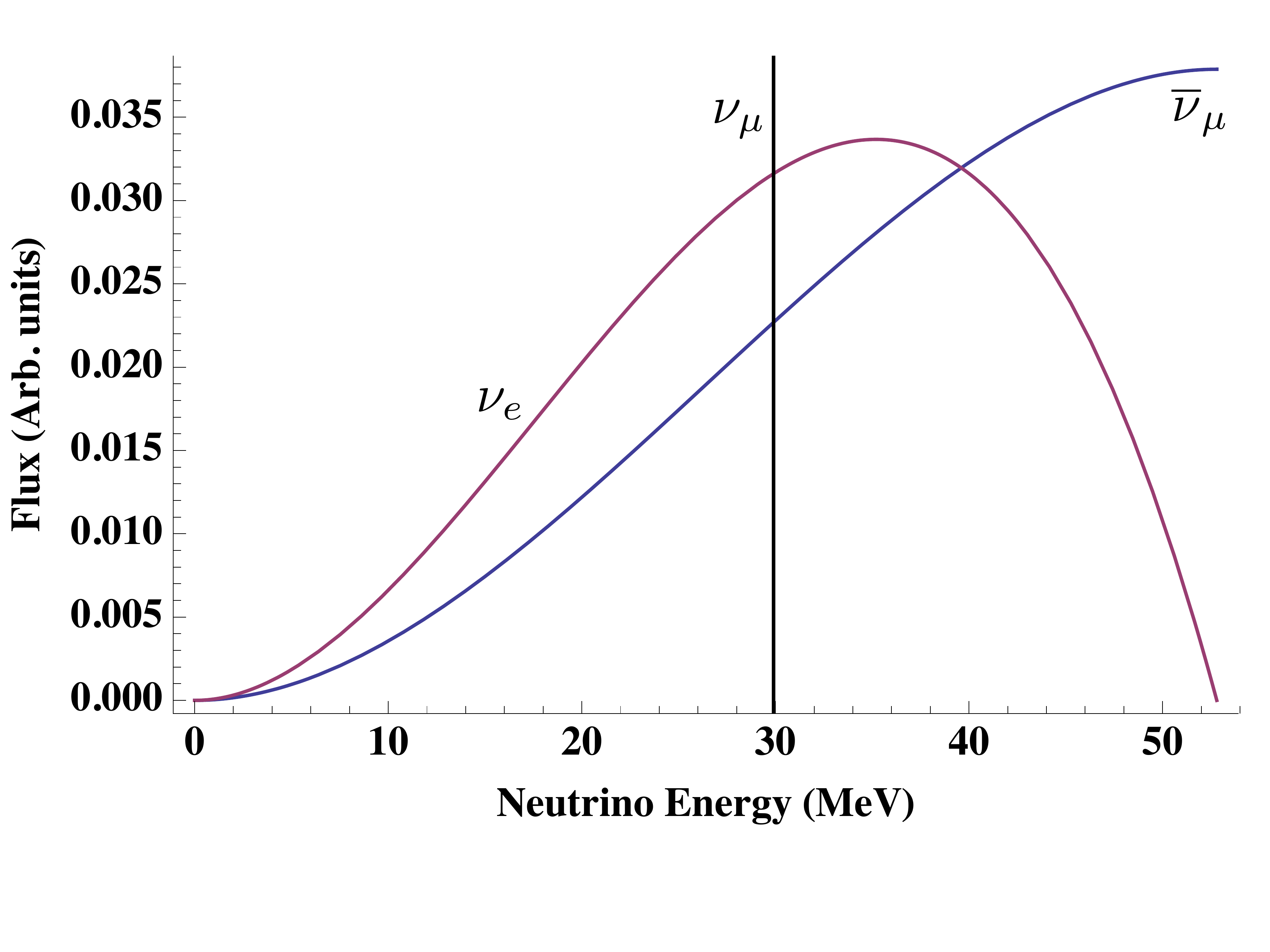}
} \end{center}
\vspace{-1.2cm}
\caption{Energy distribution of neutrinos from a DAR source.
\label{flux} }
\end{figure}

In the DAR source described here, neutrino production begins with
800~MeV protons impinging on a target to produce
low energy charged pions primarily through the $\Delta$ resonance decay.
The pion decay chain $\pi^+ \rightarrow \mu^{+}\nu_{\mu}$, $\mu^+\rightarrow e^{+}\bar{\nu}_{\mu}\nu_{e}$, produces the neutrino flux shown in Fig.~\ref{flux}. The $\bar \nu_e$ content that arises from $\pi^-$ production and subsequent decay-in-flight (DIF) is well below $10^{-3}$~\cite{KARMEN, LSND} due to $\pi^-$ capture. These features provide an ideal source for neutrino appearance
searches~\cite{LSND, KARMEN, multi, DAEdALUS}, active-to-sterile searches
relying on the charged current interaction~\cite{nuedis}, and an active-to-sterile neutral current search~\cite{garvey} as discussed in this paper.

\subsection{Targeting to allow multiple baselines}
A high intensity source of 800~MeV protons is being
developed by the DAE$\delta$ALUS collaboration~\cite{DAEdALUS}. This design
utilizes cyclotron-based accelerators~\cite{Luciano, Alonso}
installed at three sites near a very large water- or oil-based detector. 
The experiment described here could use one of these DAE$\delta$ALUS cyclotrons combined with a dual-target configuration as a neutrino source.

In a baseline scenario, the cyclotron-based beam will be diverted between the two targets so that no target receives more than 1~MW average power. The beam will be directed at 90$^\circ$ with respect to the detector, so as to minimize DIF backgrounds. Notably, a multi-target design can also be exploited for a charged current neutrino interaction oscillation measurement with a common detector and multiple baselines. The main technical issue in the two-target cyclotron design is maintaining a good vacuum in the two-prong extraction line. The
beam will be ``painted'' across the face of each target in order to
prevent hot spots in the graphite, an effect which will dominate the $\pm25$~cm uncertainty on the experimental $L$ from each neutrino source. The targets will
be arranged in a row enveloped within a single iron shield, with the detector
located 20~m downstream of the near target and 40~m downstream of the far target. This configuration has been found to provide the best overall sensitivity to the LSND allowed region.  

The analysis below exploits the $L$ dependence of neutrino oscillations. Therefore, the flux of protons on each target must be well
understood in time; standard proton beam monitors allow a 0.5\% measurement precision. The absolute neutrino flux is less important,
as sensitivity to the oscillation signal depends on relative detected rates at the various distances. The systematic uncertainty associated with the flux normalization is 10\% if there is no large water or oil detector available and 1.1\% if such a detector does exist~\cite{multi}. A high statistics $\nu$-electron scattering measurement at a large water detector provides a precise determination of the flux normalization.

\section{Detecting Coherent Neutrino Scattering}
\label{coherentintro}
Coherent neutrino-nucleus scattering, in which an incoming neutrino scatters off an entire nucleus via neutral current $Z$ exchange \cite{Freedman}, has never been observed despite its well predicted and comparatively large standard model cross section. The coherent scattering cross section is 
\begin{equation}
{{d\sigma}\over{dT}} = {{G_F^2}\over {4\pi}} Q_W^2 M \left(1-
{{MT}\over{2E_\nu^2}}\right) F(Q^2)^2~, \label{coherent}
\end{equation}
where $G_F$ is the Fermi constant; $Q_W$ is the weak
charge [$Q_W=N-(1-4~\mathrm{sin}^{2}\theta_{W})Z$, with $N$, $Z$, and $\theta_{W}$ as the number of neutrons, number of protons, and weak mixing angle, respectively]; $M$ is the nuclear target mass; $T$ is the nuclear recoil energy; and $E_\nu$ is the incoming neutrino energy. The $\sim$5\% cross section uncertainty, the actual value depending on the particular nuclear target employed, is dominated by the form factor~\cite{Horowitz:2003cz}. 

Coherent neutrino scattering is relevant for the understanding of type II supernova evolution and the future description of terrestrial supernova neutrino spectra. Measuring the cross section of the process also provides sensitivity to non-standard neutrino interactions (NSI) and a $\sin^2\theta_W$ measurement at low $Q$~\cite{scholberg}. Cross section measurements as a function of energy on multiple nuclear targets can allow the cross section dependence on NSI and $\theta_{W}$ to be isolated and understood. As demonstrated here, neutrino oscillations can also be cleanly probed using coherent scattering. 

The difficulty of coherent neutrino scattering detection arises from the extremely low energy of the nuclear recoil signature. For example, a 20~MeV neutrino produces a maximum recoil energy of about 21~keV when scattering on argon. Both a CDMS-style germanium detector~\cite{CDMS} and a single phase liquid argon detector, such as the one proposed for the CLEAR experiment~\cite{Scholberg:2009ha}, are considered in this paper for detecting these low energy events. Other dark matter style detector technologies, especially those with ultra-low energy thresholds, can be effective for studying coherent neutrino scattering as well.

\subsection{Experimental Setup}
\label{experimentsection}
The envisioned experimental setup is consistent with the current DAE$\delta$ALUS accelerator proposal and follows a realistic detector design. A single DAE$\delta$ALUS cyclotron will produce $4\times10^{22}$ $\nu$/flavor/year running with a duty cycle between 13\% and 20\% \cite{Luciano, DAEdALUS}. A duty cycle of 13\% and a physics run exposure of five total years are assumed here. With baselines of 20~m and 40~m, the beam time exposure distribution at the two baselines is optimal in a $1:4$ ratio: one cycle to near (20~m), four cycles to far (40~m). Instantaneous cycling between targets is important for target cooling and removes systematics between near and far baselines associated with detector changes over time. The accelerator and detector location is envisioned inside an adit leading into a sharp 300~ft rise at the Sanford Research Facility at Homestake, in South Dakota. The neutrino flux normalization uncertainty at each baseline is conservatively expected at 1.5\%. We assume the flux has been constrained to this level by an independent measurement of $\nu$-electron scattering with a large water-based Cerenkov detector also assumed to be in operation at Sanford Labs. The 1.5\% uncertainty estimate takes into consideration the theoretical uncertainty in the $\nu$-electron scattering cross section and the statistics achievable with a large water detector. The flux normalization correlation coefficient between the near and far baselines is conservatively set to 0.99, its deviation from unity being dominated by differences between the two beam dumps. An uncorrelated systematic uncertainty of 0.5\% at each baseline, is also included. The general experimental assumptions can be seen in Table~\ref{general_assumptions}.

\begin{table}[t]
  \begin{center}
    {\footnotesize
      \begin{tabular}{|c|c|}
      \hline
        $\nu$ source& $4\times10^{22}$ $\nu$/flavor/year\\ 
        Duty factor & 13\%\\ 
        Baseline correlation & 0.99 \\
        $\nu$ flux norm. uncertainty & 1.5\%   \\
        Uncorr. sys. uncertainty & 0.5\%   \\
        Distances from $\nu$ source & 20~m, 40~m\\       
        Exposure & 5~years: 1 near, 4 far \\ 
        Depth & 300 ft \\ \hline 
      \end{tabular} 
      \caption{The experimental configuration assumptions.}\label{general_assumptions}
}
\end{center}
\end{table}

We also consider a ``dedicated" physics run scenario in which the duty factor is raised from 13\% to 50\% for all five years. With the instantaneous power achievable remaining constant, this change leads to an average power increase of a factor of 4. Steady-state and beam-related backgrounds also increase by this factor in a dedicated scenario. The nominal duty factor of 13\% is driven by the requirement that the various DAEdALUS accelerator baseline beam windows do not overlap in time. A dedicated scenario is possible in consideration of maintaining sufficient target cooling and the phased DAE$\delta$ALUS deployment timeline. The timeline calls for a cyclotron or set of cyclotrons installed exclusively at a single ``near" baseline, close to a large water detector, for at least five years~\cite{DAEdALUS}. With a 13\% duty factor only required when all baselines have operational accelerators, a longer duty factor and higher average power seems possible in DAE$\delta$ALUS single-baseline-only operation. Note that although only two targets are required for the experimental design described here, supplementing the beamline with more targets can ensure optimal use of beam time in consideration of cooling requirements and ultimately increase neutrino oscillation sensitivity. 

\subsubsection{Germanium detector -- signal and backgrounds}
A low-threshold germanium-based detector, such as CDMS, measures phonons and ionization from electronic and nuclear recoils~\cite{cdmscite1}. A CDMS detector consists of a large germanium crystal ($0.25-1$~kg) operated at cryogenic temperatures ($\sim$$100$~mK) with thousands of superconducting transition-edge sensors (TESs) photolithographically patterned on the top and bottom surfaces.  The TESs are wired in parallel to form four readout channels on each surface, which measure phonons created in particle interactions. The particle-induced ionization is also measured by electrodes on the crystal surface. The ratio of the energy in these two channels is a powerful discriminator between nuclear and electronic recoils.

A 100~kg active mass of germanium is considered for the experiment described here, similar to proposed dark matter searches~\cite{CDMSSnoLab}. The detection efficiency above a 10~keV threshold is set to 0.67 with a 3\% energy resolution near the threshold. These assumptions are reasonably conservative and consistent with future expectations~\cite{cdmsenergyres,iZIPdiscrimination}. 

Two classes of background events are considered for a germanium detector:
\begin{enumerate}
\item \emph{Misidentified electronic recoils} - Electronic recoils can be produced by photons and beta particles interacting with the active detection medium. Misidentification of such events is particularly problematic near the detector surfaces, where the collection of electron-hole pairs is suppressed and discrimination is less effective. Existing experiments have demonstrated an electronic recoil misidentification rate of less than 1~event per 100~kg$\cdot$days exposure~\cite{CDMS}. Upgrades to detector design are expected to improve discrimination by a factor of $10^4$~\cite{iZIPdiscrimination}. The assumed rate of radiogenic background detection ($\sim$2~events/year) is negligible. 

\item \emph{Cosmogenic neutrons} - Single scatter neutrons can produce a signal identical to a coherent neutrino scattering event, and the rate of these events would be significant at a shallow site. As a point of reference for surface experiments, the CDMS experiment located at the Stanford Underground Facility with 16~m.w.e. of overburden measured a neutron background of 0.67~events/(kg$\cdot$day)~\cite{cdmsSUF}. This figure could be significantly reduced with additional active and passive shielding and the larger overburden envisioned for the DAE$\delta$ALUS site. A cosmogenic-induced background of 0.1~detected~events/(10 kg$\cdot$day), after correcting for efficiency and during beam-on, is assumed. This value is considered a design goal and can be met with a 300~ft overburden and modest active and passive shielding.
\end{enumerate}

In this study, the estimated radiogenic and cosmogenic background rates are distributed evenly across the germanium nuclear recoil energy range considered, 10~keV to 100~keV.

\subsubsection{Liquid argon detector -- signal and backgrounds}
A single phase liquid argon detector can be used to detect the scintillation light created by WIMP- or coherent neutrino-induced nuclear recoils. Such detectors employ a large, homogeneous liquid argon volume surrounded by photomultiplier tubes (PMTs). Inner detector surfaces as well as the PMTs themselves are usually covered in a wavelength shifting substance which converts the 128~nm scintillation light into the visible spectrum for detection. 

A 456~kg active mass of liquid argon with a flat efficiency of 0.50 above a 30~keV energy threshold is considered for the experiment described here. The detection volume and efficiency are consistent with the proposed CLEAR design~\cite{Scholberg:2009ha}. An 18\% energy resolution near threshold is used, assuming resolution slightly worse than what would be expected from photoelectron Poisson statistics~\cite{kryptoncal}, 6~photoelectrons/keVee light collection, and a quenching factor of 0.25~\cite{gastlerAr}. 

There are three primary sources of background that are considered for a single phase liquid argon detector: 

\begin{enumerate}
\item \emph{Cosmogenic neutrons} - The muon-induced neutron background is contingent on the geometry of the site, overburden, and active/passive detector shielding. Muon events and muon-induced neutrons can be vetoed with high efficiency and low detector dead time in a liquid argon detector near the surface~\cite{Scholberg:2009ha}. The target design cosmogenic background is 0.1~detected~events/(10~kg$\cdot$day).
\item \emph{$^{39}$Ar contamination} - $^{39}$Ar is a naturally occurring radioisotope with an isotopic abundance of $^{39}$Ar/Ar $= 8 \times 10^{-16}$, corresponding to a specific activity of 1.01~Bq/kg~\cite{39ArActivity}. The isotope is a beta emitter with an energy endpoint of 565~keV.

Pulse-shape discrimination (PSD) can be used to separate the $^{39}$Ar-induced electronic recoils from the nuclear recoils produced in WIMP and coherent neutrino scattering events~\cite{BoulayAndHime}. The electronic recoil contamination (ERC) of the nuclear recoils decreases exponentially as the number of photoelectrons detected increases. Reference~\cite{DEAP1ERC} measures the ERC for PSD in the single phase liquid argon DEAP-1 detector (4.85 photoelectrons per keVee) and also provides a ``theoretical" Monte Carlo estimate of the ERC attainable for an ideal detector with 4$\pi$ PMT coverage and 6~photoelectrons/keVee. Both scenarios correspond to a 50\% efficiency for nuclear recoil detection in the fiducial volume. Note that, according to Ref.~\cite{himemicroclean}, the MicroCLEAN experiment has achieved 6~photoelectrons/keVee sensitivity. The abundant $^{39}$Ar background could further be alleviated with the use of depleted argon from underground sources, which has an isotopic abundance of $^{39}$Ar that is $<$5\% of natural argon at the surface~\cite{DepletedAr}. Figure~\ref{fig:Ar} shows the rate of $^{39}$Ar after PSD with 13\% on-time, for two assumptions of ERC reported in Ref.~\cite{DEAP1ERC}. The theoretical ERC with non-depleted liquid argon is employed for this study. 

\begin{figure}
\begin{center}
\includegraphics[scale=0.4]{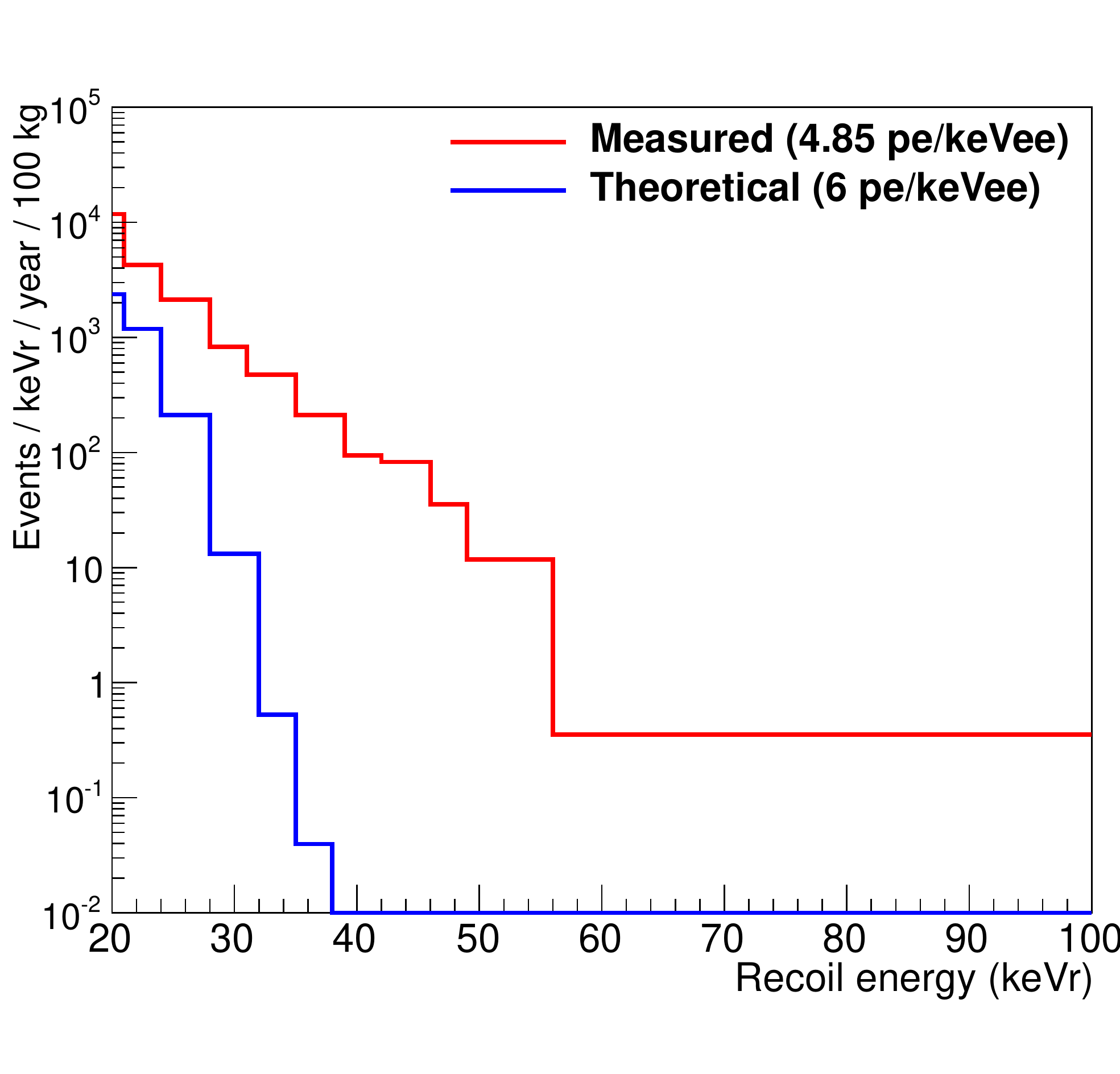}
\vspace{-.7cm}
\caption{The expected $^{39}$Ar background energy spectrum under two sets of assumptions. The line labeled ``Measured" corresponds to an ERC that was obtained in a detector with 4.85~photoelectrons/keVee. The line labeled ``Theoretical" is the ERC simulated in an ideal detector with 6~photoelectrons/keVee and represents the background ERC used for this study.  Both lines correspond to a 50\% efficiency for detecting nuclear recoils in the fiducial volume.\label{fig:Ar}}
\end{center}
\end{figure}

\item \emph{Surface contamination} -  Radioactive impurities on the detector surfaces can decay and contribute to the background. These surface backgrounds have been measured in the DEAP-1 detector and were found to have an activity of $1.3 \times 10^{-4}$~Bq~\cite{DEAP1bkgd}. Depending on the origin of these events, the scaling and resulting background prediction can differ. If the events are due exclusively to $^{210}$Pb surface contamination, the DEAP-1 figure can be scaled by detector surface area to yield $1.3 \times 10^4$~events/year. However, this rate may be substantially reduced by the use of cleaner materials, scrubbing of the surface, and fiducialization. A surface background contamination of 100~detected~events/year is assumed here and can be considered a design goal. 
\end{enumerate}

The 30~keV energy threshold employed here is larger than the oft-chosen 20~keV threshold in single phase liquid argon detectors in order to mitigate the steeply falling $^{39}$Ar contamination. If $^{39}$Ar discrimination improves in a future design, adjusting the threshold to (e.g.) 20~keV would allow a 60\% larger signal sample. In this study, the estimated surface and cosmogenic background rates are distributed evenly across the argon nucleus recoil energy range considered, 30~keV to 200~keV.

One additional possibility that would significantly reduce the non-beam-related background would be to use a pulsed source of neutrinos, such as at the Spallation Neutron Source (SNS). The SNS produces protons in very short bunches of $<$750~nsec at a rate of about 60~Hz, so that the time window for expected signal events is a small fraction of the total running time. Combining a pulsed DAR beam structure with a liquid argon detector was previously proposed by the CLEAR experiment~\cite{Scholberg:2009ha}, allowing them to claim an additional rejection of $6 \times 10^{-4}$ for steady-state, non-beam-related backgrounds using a timing cut. Although the DAE$\delta$ALUS proposal does not include this timing structure, the experimental concept described here could be employed at other facilities.

\begin{figure}[h]
\begin{center}
\includegraphics[scale=0.4]{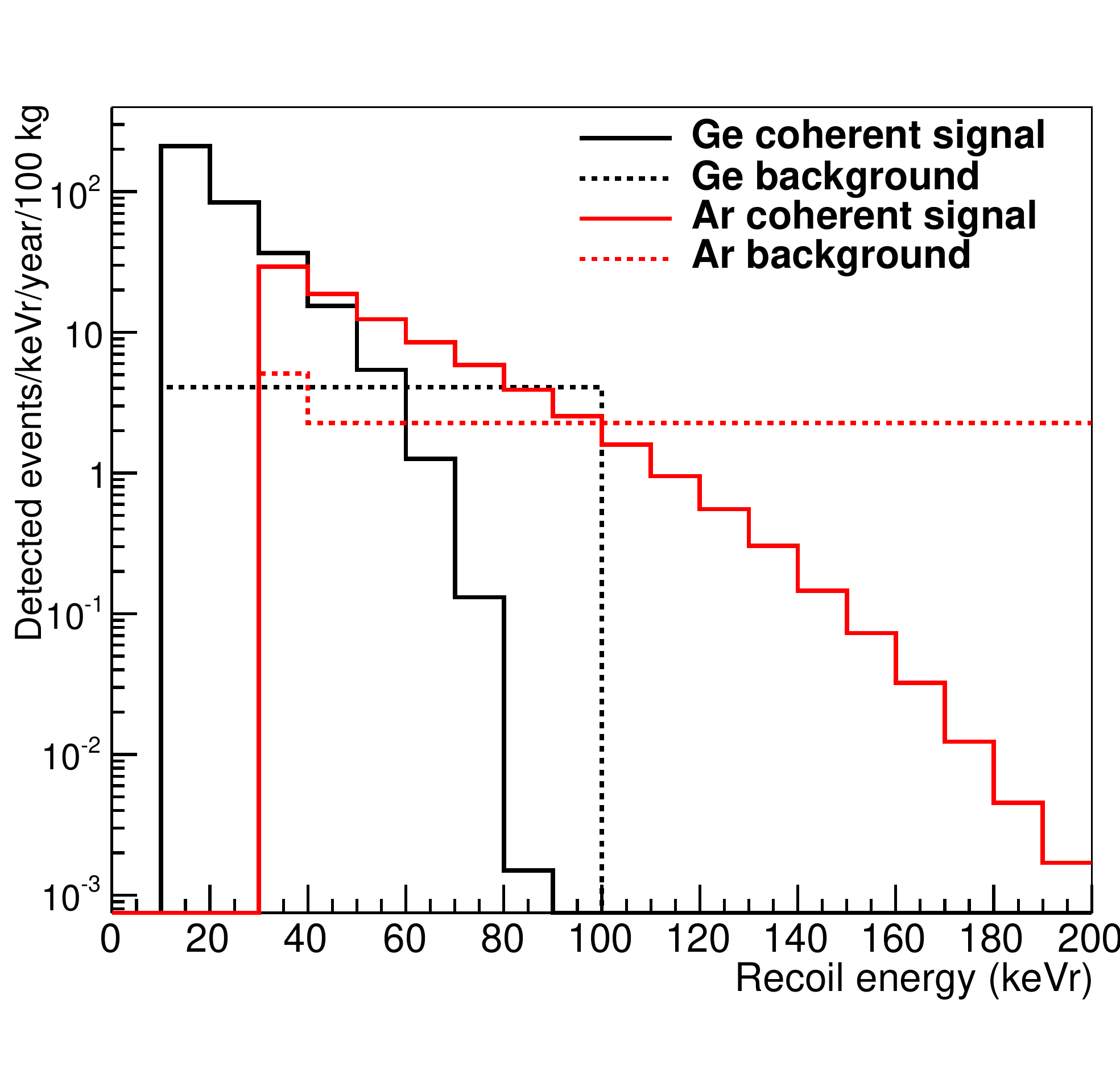}
\vspace{-.7cm}
\caption{The expected non-oscillated signal and total background rates at a 20~m baseline for the two detector technologies considered in the baseline physics run scenario. The rates correspond to what is expected for one full year of near-target-only operation at 13\% duty factor. }
\label{fig:rateplot}
\end{center}
\end{figure}

The detector-specific assumptions are summarized in Table~\ref{target_assumptions} and the expected signal and background rates are shown in Fig.~\ref{fig:rateplot}.

\begin{table}[t]
\label{base_opti}
  \begin{center}
    {\footnotesize
      \begin{tabular}{|c|c|c|} \hline 
        &$^{76}$Ge & $^{40}$Ar \\ \hline
        Active mass & 100~kg & 456~kg \\
        Efficiency & 0.67 (flat)   & 0.50 (flat) \\
        Threshold &10 keV &30 keV \\
        $\frac{\mathrm{\Delta E}}{\mathrm{E}}$ at threshold&3\%& 18\%\\        
        Radiogenic background&2/year & See text  \\
        Cosmogenic background&0.1/(10 kg$\cdot$day) & 0.1/(10 kg$\cdot$day)  \\
        Beam-related background&0/year & 0/year  \\ \hline
      \end{tabular} 
      \caption{The assumptions relevant for the specific detector technologies considered.}\label{target_assumptions}
}
\end{center}
\end{table}

\subsubsection{Neutron flux from the source}
DAR sources produce a large flux of neutrons, arising from spallation reactions of protons with the beam dump material. For the DAR source considered here, the neutrons have energies up to 800~MeV. In a 1~MW beam, the neutron production rate is $\sim$$10^{16}$/s and attenuation lengths may be as high as tens of centimeters. Single scatter neutrons can produce elastic recoils in the detector volume that are indistinguishable from coherent neutrino scattering on an event-by-event basis. Moreover, because the neutron flux is attenuated by matter, underestimating the neutron background in the detector could mimic a deviation from the $1/r^2$-dependence of the neutrino flux, similar to what is expected for neutrino disappearance. It is therefore essential to locate the detector far enough from the source that the beam-related neutron flux is negligible.
 
A precise estimate of the neutron flux would require detailed knowledge of the experimental site, beam configuration, and shielding. The neutron flux is estimated with a Monte Carlo simulation of the experimental geometry consistent with the DAE$\delta$ALUS proposal \cite{DAEdALUS} and several simplifying assumptions.  Instead of simulating the passage of neutrons through the beam dump shielding, we simply assume that a cubic shield with sides of length 6~m is sufficient to reduce the escaping neutron flux to a level consistent with safety regulations. Also, we assume that this cube of shielding is adjacent to a rock (SiO$_2$) cliff. The maximum permissible annual dose for workers in a restricted area with a neutron beam is 100~mRem~\cite{RadiationLimits}. The neutron flux escaping the shielding is set to a rate equivalent to an exposure of 100~mRem in 40~hours.

Using a Geant4-based simulation~\cite{geant}, neutrons are injected at the edge of the shielding cube. The neutrons are simulated in energy bins from 0-30~MeV. The flux is tallied at 20~cm intervals into the rock cliff, and the fluxes beyond 1~m into the cliff are fit to the functional form
\begin{equation}
\Phi(z) = \frac{Ae^{-z/\lambda}}{z^2}~,
\end{equation}
where $A$ and $\lambda$ are fit parameters, and $z$ is the distance from each flux tally point to the DAR source. The neutron fluxes are in reasonable agreement with this functional form. The fit function is then used to extrapolate the flux to a full year of running and larger distances from the source. A simulation is also employed to estimate the fraction of incident neutrons that produce single-scatter nuclear recoils in the detection volume. Less than 0.2 beam-related events are expected per year for a 456~kg liquid argon detector at a 12~m baseline. The beam-related background at 20~m from the source, the shortest relevant detector baseline considered here, is therefore assumed to be negligible.

\section{Measurement Strategy and Sensitivity}
\label{sensitivitysection}
\subsection{Overall strategy}
Neutrino oscillations depend upon neutrino energy and distance traveled. Since the neutrino energy cannot be reconstructed precisely with the coherent interaction,
our sensitivity to the oscillatory behavior arises mainly from $L$, a value which is well determined by the location of the target being used at any given time and its distance to the common detector. In the case that a disappearance signal is detected, the target exposure priorities for the two baselines can be optimized to maximize sensitivity.


The purely neutral current experiment described is sensitive to the effective disappearance of all three types of neutrinos present in the beam, $\nu_\mu$, $\bar{\nu}_\mu$, and $\nu_e$, into $\nu_s$. We assume this disappearance can be approximated by a two-neutrino oscillation driven by a $\Delta m^2$ in the LSND allowed region, and that the oscillation probability under the approximation is the same for neutrinos and anti-neutrinos. The baselines for the experiment, 20~m and 40~m, have been chosen in order to provide the best sensitivity to the LSND allowed parameter space, given the neutrino energy spectrum of each flavor in the beam. The experiment described here provides indirect sensitivity to the LSND allowed parameter space by simultaneously measuring terms describing the amplitude of active neutrino mixing to a sterile neutrino: $4|U_{e4}|^2|U_{s4}|^2$ in the case of $\nu_e$ in the beam, and $4|U_{\mu4}|^2|U_{s4}|^2$ in the case of $\nu_{\mu}$ and $\bar{\nu}_{\mu}$ in the beam. These terms are then translated to the appearance amplitude measured by LSND, $\sin^22\theta_{\mu e}=4|U_{e4}|^2|U_{\mu4}|^2$. Sensitivity to $\sin^22\theta_{\mu e}$, along with simultaneous sensitivity to $\sin^22\theta_{ee}=4|U_{e4}|^2(1-|U_{e4}|^2)$ and $\sin^22\theta_{\mu\mu}=4|U_{\mu4}|^2(1-|U_{\mu4}|^2)$, are considered the figures of merit here, as they can be easily compared to existing charged current appearance and disappearance measurements. Of course, distinguishing between $\sin^22\theta_{ee}$ and $\sin^22\theta_{\mu\mu}$ in the case of an observed disappearance is not possible in a flavor-blind experiment. Therefore, we rely on marginalizing over the full parameter space of $|U_{\mu4}|$ and $|U_{e4}|$ explored, in the most conservative case possible, when drawing sensitivity contours for each case. 

The sensitivity to any particular set of oscillation parameters is obtained by simultaneously fitting the expected flavor-summed coherent signal events as a function of recoil energy at the near and far baselines. The events at each baseline are distributed among bins of nuclear recoil energy (1~bin/10~keV); however, the sensitivity results are largely insensitive to the number of recoil energy bins used in the comparison.

\subsection{Sensitivities}
The signal predictions are evaluated for each set of oscillation parameters, $\Delta m^2_{41}\equiv\Delta m^2$, $|U_{\mu4}|$, and $|U_{e4}|$. A $\chi^2$ is calculated by comparing the oscillations-predicted spectra, including backgrounds, to the no-oscillations prediction.

The $\chi^2$ is constructed as
\begin{equation}
\chi^2 = \sum_{i,j = 1}^{N_{\mathrm{bins}}} (P_i-N_i)(P_j-N_j)M^{-1}_{ij}~,
\end{equation}
where $i$ and $j$ denote the energy bins at the near and far baselines, respectively; $P_{i}$ is the oscillations-predicted event spectrum as a function of $N_{\mathrm{bins}}=1,...,10,11,...,20$ bins, corresponding to (e.g.) 10 energy bins for the two baselines appended side by side; $N_i$ is the corresponding no-oscillations spectrum; and $M^{-1}_{ij}$ is the inverse covariance matrix including statistical and systematic uncertainties and normalization systematic correlations between the two baselines and different recoil energy bins. Note that the background contributions to $P_i$ and $N_i$ cancel. The background-contributed statistical uncertainty, however, is accounted for in $M_{ij}$. The background contribution can be measured with high statistics during beam-off cycles, and so systematic uncertainties associated with background are small relative to statistical uncertainties.

The oscillations-predicted spectra, $P_{i}$, are obtained by summing over all neutrino flavors predicted in each recoil energy bin of the unoscillated spectrum, and reweighting each neutrino according to its flavor $\alpha=e,\mu$ by the following ``active'' survival probability
\begin{eqnarray}
P(\nu_{\alpha}\rightarrow\nu_{\mathrm{active}}) &=& 1 - P(\nu_{\alpha}\rightarrow\nu_s) \nonumber \\
&=& 1 - \sin^22\theta_{\alpha s}\sin^2(1.27\Delta m^2 L/E)~,
\end{eqnarray}
where $\nu_{\mathrm{active}}$ can be any active state including $\nu_{\alpha}$, and $\sin^22\theta_{\alpha s}=4|U_{\alpha4}|^2|U_{s4}|^2$. By unitarity assumptions, $|U_{s4}|^2$ is a function of $\sum_{\alpha=e,\mu,\tau}|U_{\alpha4}|^2$,
\begin{equation}
|U_{s4}|^2=1-\sum_{\alpha=e,\mu,\tau}|U_{\alpha 4}|^2~.
\end{equation}
During the fit, we vary $|U_{e4}|$, $|U_{\mu4}|$, and $\Delta m^2$. For simplicity, however, we assume $|U_{\tau 4}|=0$. Note that a non-zero $|U_{\tau 4}|$ would increase the active survival probability for any given $|U_{e4}|$ and $|U_{\mu4}|$, and would therefore make this search slightly less sensitive to oscillations in terms of $\sin^22\theta_{\mu e}$. On the other hand, if non-zero $|U_{e4}|$ and $|U_{\mu4}|$ were to be established independently by other short baseline experiments, the type of neutral current search outlined in this paper may offer sensitivity to $U_{\tau4}$, depending on the sizes of $|U_{e4}|$, $|U_{\mu4}|$ and $|U_{\tau4}|$.

\begin{figure}[h]
\begin{center}
\includegraphics[scale=0.4]{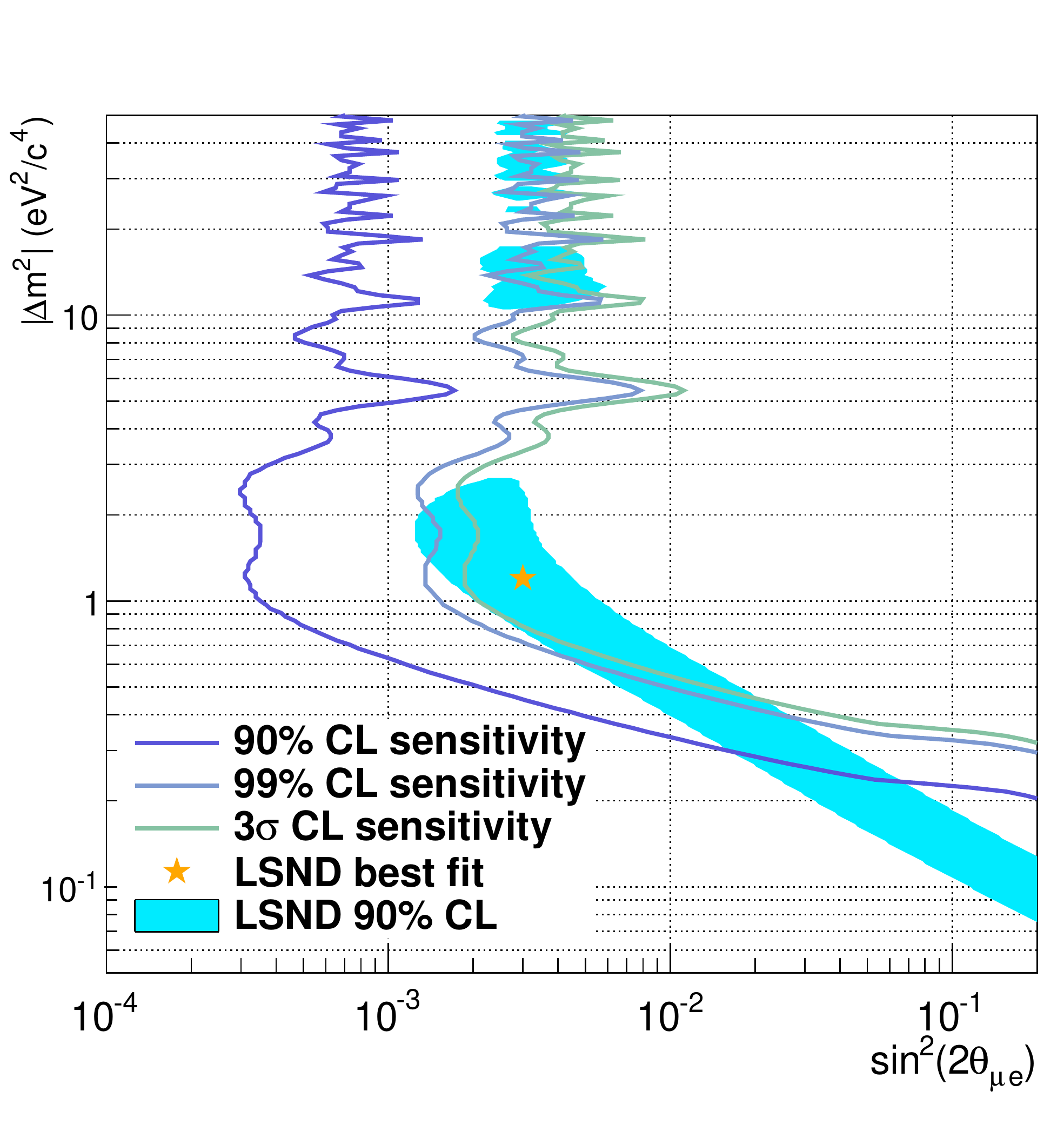}
\vspace{-.7cm}
\caption{Sensitivity to the LSND 90\% CL allowed parameter space with a germanium-based detector under the baseline physics run scenario.}
\label{fig:mue_ge76}
\end{center}
\end{figure}

\begin{figure}[h]
\begin{center}
\includegraphics[scale=0.4]{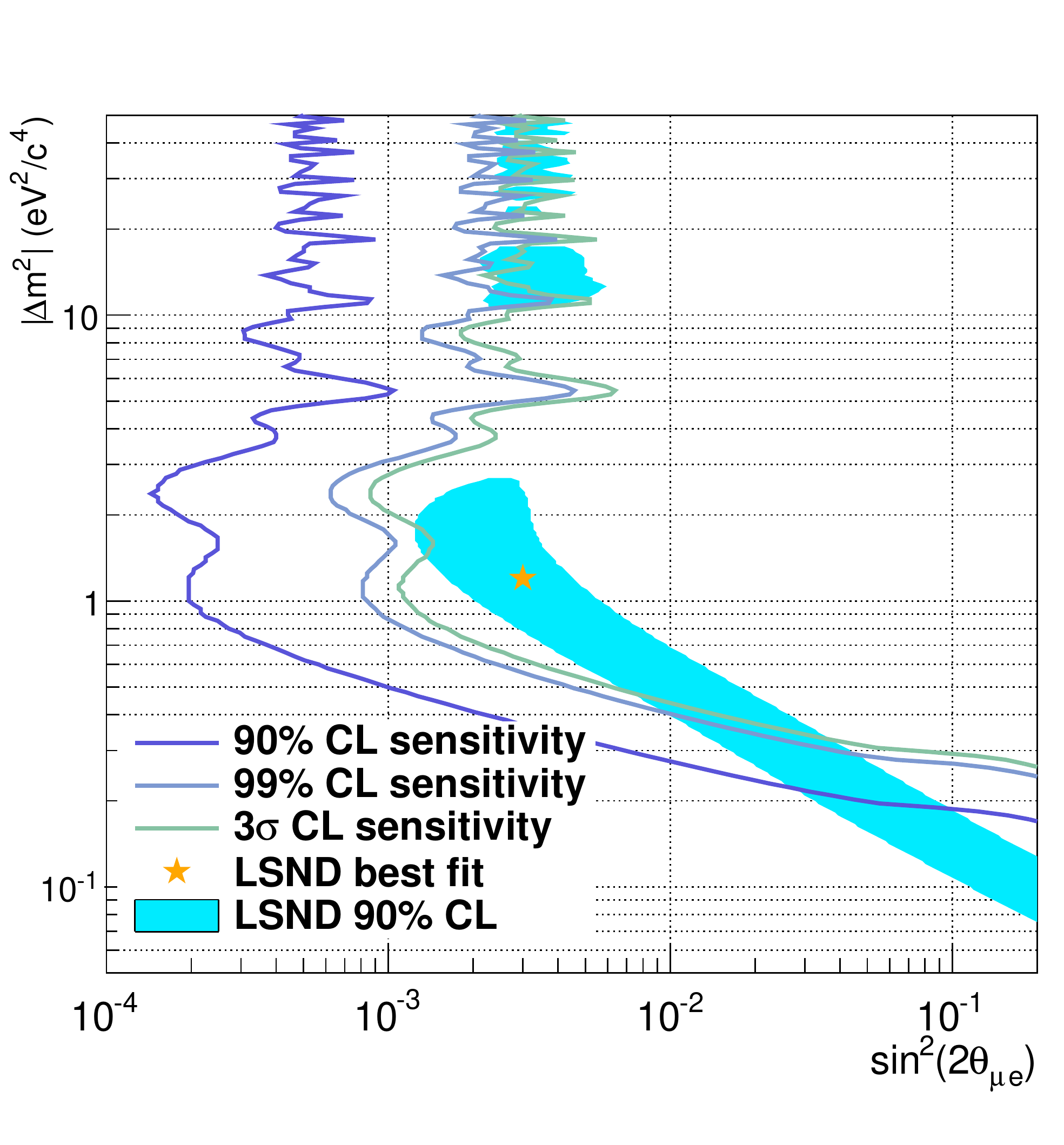}
\vspace{-.7cm}
\caption{Sensitivity to the LSND 90\% CL allowed parameter space with a germanium-based detector under the dedicated physics run scenario.}
\label{fig:mue_20years_ge76}
\end{center}
\end{figure}

\begin{figure}[h]
\begin{center}
\includegraphics[scale=0.4]{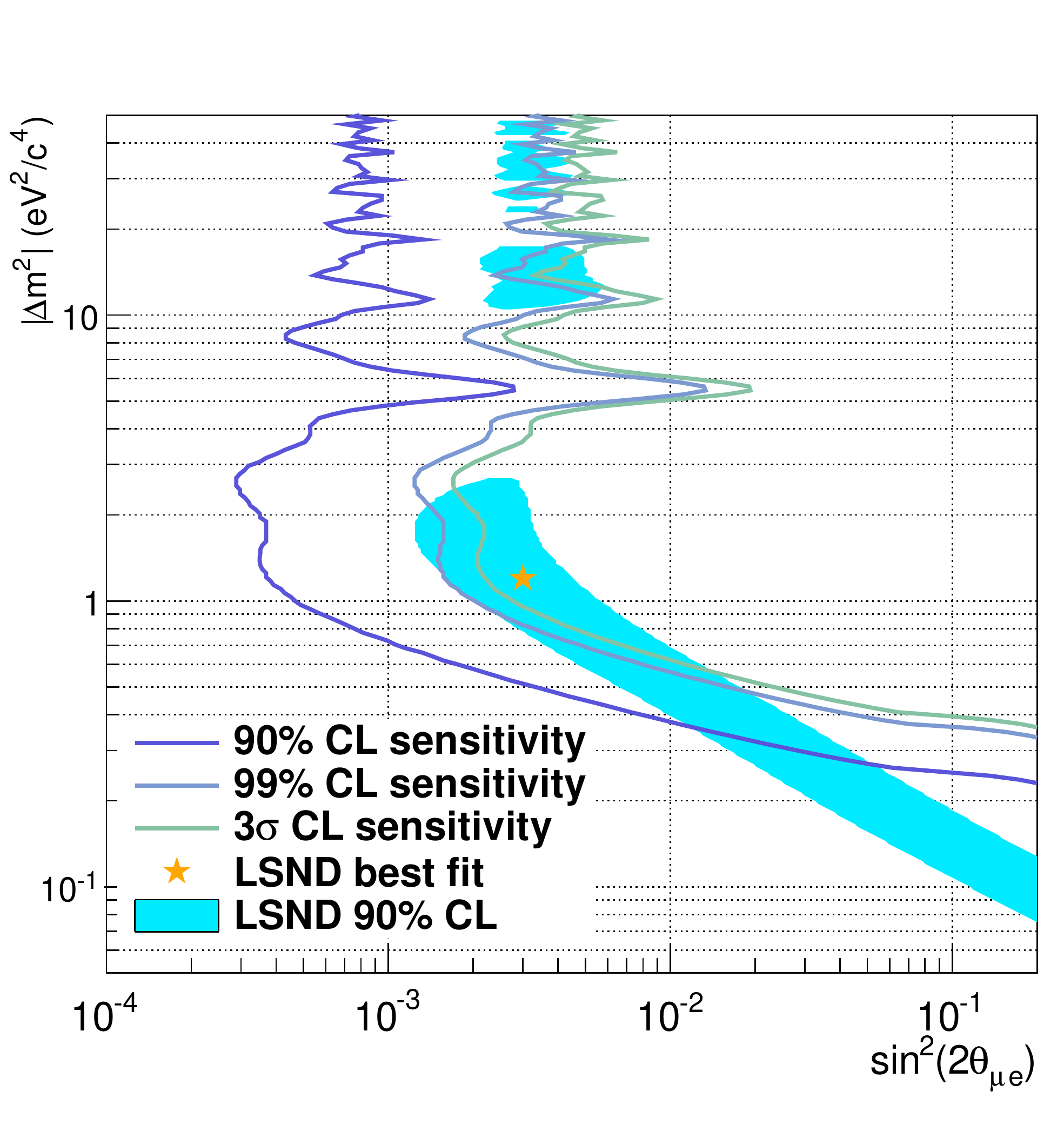}
\vspace{-.7cm}
\caption{Sensitivity to the LSND 90\% CL allowed parameter space with an argon-based detector under the baseline physics run scenario. }\label{fig:mue_ar40}
\end{center}
\end{figure}

Figures~\ref{fig:mue_ge76}, \ref{fig:mue_20years_ge76} and \ref{fig:mue_ar40} show the expected sensitivity to the LSND allowed region with a germanium detector in the baseline and dedicated physics run scenarios and an argon detector in the baseline scenario, respectively. In obtaining the sensitivity curves, the 3D search grid is reduced from ($\Delta m^2$, $|U_{e4}|^2$, $|U_{\mu4}|^2$) to a 2D space of $\Delta m^2$ and $\sin^22\theta_{\mu e} =4|U_{e4}|^2|U_{\mu4}|^2$. Note that a non-zero $\sin^22\theta_{\mu e}$ requires both $\nu_e$ and $\nu_{\mu}$ disappearance.

The $\sin^22\theta_{\mu e}$ sensitivity curves are obtained using a raster scan in $\Delta m^2$ space. That is, each curve maps out the maximum $\sin^22\theta_{\mu e}=4|U_{e4}|^2|U_{\mu4}|^2$ which satisfies $\chi^2\le\Delta\chi^2_{cut}$ at a given confidence level, for each point in $\Delta m^2$.  The 90\%, 99\%, and 3$\sigma$ confidence level curves shown in this paper correspond to $\Delta\chi^2_{cut}=$1.64, 6.63, and 9.00 for a one degree of freedom, one-sided raster scan (90\%), and a one degree of freedom, two-sided raster scan (99\% and 3$\sigma$), respectively.

\begin{figure}[h]
\begin{center}
\includegraphics[scale=0.4]{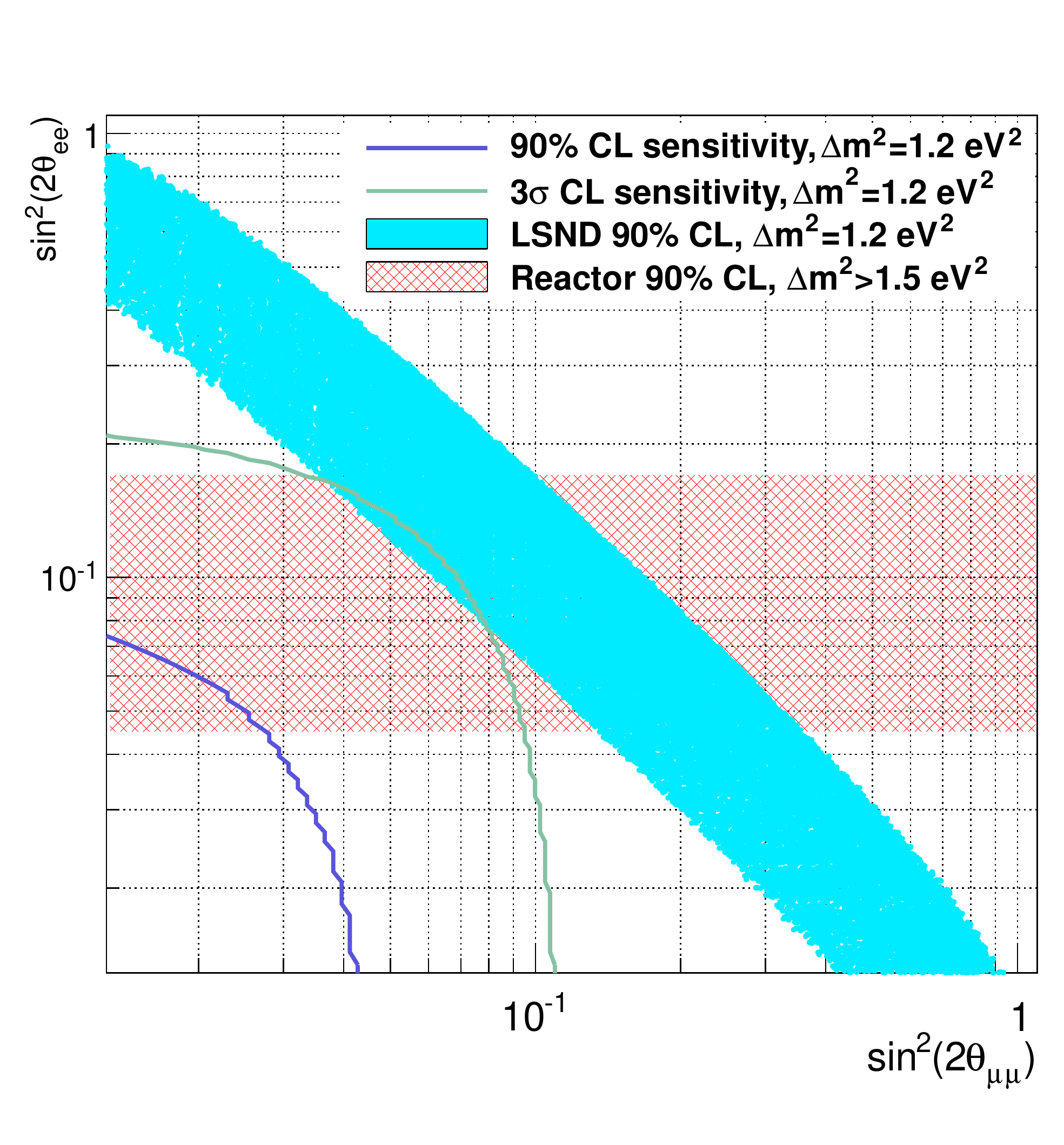}
\vspace{-.7cm}
\caption{Sensitivity to disappearance amplitudes accessible with charged current searches, assuming the LSND best-fit $\Delta m^2=1.2~\rm{eV}^2$. The sensitivity corresponds to a germanium-based detector under the baseline physics run scenario. The LSND band represents the 90\% CL allowed values of $\sin^22\theta_{\mu e}$ at $\Delta m^2=1.2~\rm{eV}^2$. ``Reactor" refers to the result reported in Ref.~\cite{Lassere} and indicates the range of $\sin^22\theta_{ee}$ values preferred by a joint fit to reactor and gallium experiment calibration measurements. The reactor result is nearly independent of $\Delta m^2$, for $\Delta m^2$ values near and above 1.5~eV$^2$.}
\label{fig:ee_mumu_ge76}
\end{center}
\end{figure}

Figure~\ref{fig:ee_mumu_ge76} shows the oscillation sensitivity for a germanium detector in terms of the disappearance amplitudes which would be accessible in charged current searches, $\sin^22\theta_{ee}=4|U_{e4}|^2(1-|U_{e4}|^2)$ and $\sin^22\theta_{\mu\mu}=4|U_{\mu4}|^2(1-|U_{\mu4}|^2)$ overlaid with the region allowed by LSND at 90\% CL, assuming the LSND best-fit $\Delta m^2=1.2~\mathrm{eV^2}$. The curves are obtained using a one-sided raster scan in $\sin^22\theta_{ee}$ with the $\Delta\chi^2_{cut}$ values defined above. The figure also shows the approximate region of $\sin^22\theta_{ee}$ values allowed at 90\% CL by fits to the reactor anomaly and gallium experiment calibration data sets in Ref.~\cite{Lassere}. The ``reactor" allowed contour is for $\Delta m^2\gtrsim$1.5~eV$^2$ and is relatively independent of $\Delta m^2$ in this region. As a reference, limits on $\sin^22\theta_{\mu\mu}$ from the MINOS neutral-current oscillation search correspond to $\sin^22\theta_{\mu\mu}<0.1$ at 90\% CL, for $\Delta m^2=1.2~\rm{eV}^2$~\cite{MinosNC}.

Figures~\ref{fig:mue_ge76} and \ref{fig:mue_ar40} show that, despite the difference in fiducial mass, the 100~kg germanium detector performs slightly better than the 456~kg liquid argon one. The difference is in part due to the difference in nuclear recoil energy threshold; 10~keV for germanium, 30~keV for argon. This emphasizes the fact that a low detector energy threshold is important for obtaining a high-statistics sample of coherent neutrino scattering events as the rate is dominated by events with very low energy recoils ($\lesssim$10~keV). 

In a baseline physics run scenario, an experiment featuring a germanium- or argon-based detector can exclude the LSND best-fit mass splitting ($\Delta m^2=1.2~\mathrm{eV^2}$) at 3.8$\sigma$ or 3.4$\sigma$, respectively. The LSND best-fit mass splitting is excluded at 4.8$\sigma$ in the dedicated, germanium-based physics run scenario considered. For sensitivity in terms of $\sin^22\theta_{ee}$ and $\sin^22\theta_{\mu\mu}$, a germanium-based experiment in the baseline scenario could exclude nearly all of the available 90\% CL LSND parameter space at the 3$\sigma$ level and large portions of the available reactor anomaly allowed region, assuming $\Delta m^2\sim 1.2~\rm{eV}^2$. 

\section{Conclusions}
This paper has described a method to search for active-to-sterile neutrino oscillations
at relatively short baselines using neutral current coherent neutrino-nucleus scattering.
Detection of such a process could definitively establish the existence of sterile
neutrinos and measure their mixing parameters.

An experiment that relies on the high statistics detection of an as-yet-undetected
process is obviously difficult. However, all of the technology required for such an experiment either exists or has been proposed with realistic assumptions. A cyclotron-based proton beam can be directed to
a set of targets, producing a low energy neutrino source with multiple baselines. This allows a measurement of the distance dependence of an
oscillation signal without moving detectors or instrumenting multiple devices. Both a germanium-based detector inspired by the CDMS design and a liquid argon detector inspired by the proposed CLEAR experiment would be effective for performing these measurements. 

Along with relevance in understanding Type II supernova evolution and supernova neutrino detection, coherent neutrino-nucleus scattering can provide sensitivity to non-standard neutrino interactions, the weak mixing angle, and, as shown in this paper, neutrino oscillations at $\Delta m^2 \sim 1$~eV$^2$. Depending on the detector technology and run scenario, the experiment described is sensitive to the LSND best-fit mass splitting at the level of 3-5$\sigma$ and can probe large regions of the LSND and reactor anomaly allowed regions. The experiment offers a pure and unique analysis of neutrino oscillations that is complementary to charged current-based appearance and disappearance searches. 

\begin{center}
{ {\bf Acknowledgments}}
\end{center}
We thank the National Science Foundation and Department of Energy for support.

\end{document}